# The Role of Occlusion: Potential Extension of the ICH E9 (R1) Addendum on Estimands and Sensitivity Analysis for Time-to-Event Oncology Studies

*Jonathan M Siegel (Bayer), Hans-Jochen Weber (Novartis), and Stefan Englert (AbbVie) & on behalf of the Pharmaceutical Industry Working Group on Estimands in Oncology*

**Abstract**

The ICH E9 (R1) Estimands Guidance[1] terminology does not completely address the conceptual needs of time-to-event estimands in the complex oncology context. We previously described how censoring and censoring mechanisms for time-to-event endpoints can be embedded into the ICH E9 (R1) Estimands Guidance teminology.[2] This second paper by the Pharmaceutical Industry Working Group on Estimands in Oncology Censoring Mechanisms Subteam discusses special issues in the oncology clinical context that may require different approaches than some other therapeutic areas as well as an extensions of the ICH E9 (R1) guidance. The concept of censoring is discussed in the broader context of occluding events, with occlusion representing any loss to further follow-up and/or removal of further collected data from analysis. Occlusion constitutes a broader concept than the estimand guidance's "intercurrent event" and "terminal event" terminology and is appropriate to describe and handle situations like withdrawal from assessments or situations where the requirements of different estimands conflict. We characterize, provide additional details, practical implications, and examples on the application of each estimands strategy for handling occluding events.

**Introduction**

This paper is motivated by discussions of the Pharmaceutical Industry Working Group on Estimands in Oncology Censoring Mechanisms Subteam and poses a potential extension of the ICH E9 (R1) addendum on estimands and sensitivity analysis in clinical trials for time-to-event oncology studies. We take a more critical look at the Estimands Guidance terminology, and highlight areas where it does not fulfil the conceptual needs of time-to-event estimands in the complex oncology context. Expanding on the concepts of noumenal and phenomenal censoring introduced in our first paper,[2] we focus on the role of occlusion, which we propose as a generalization of censoring to support the discussion of handling of events that result in not collecting or discarding subsequent data in a wider variety of ways. We discuss the impact of the clinical context of oncology, including its high mortality, heterogeneity, uncertainty as to underlying causes, complex tumor dynamics, and non-proportional hazards. We focus on the importance of goals, clinical trials as prediction; analytical and enumerative studies; the importance of understanding underlying processes; and the role of system and interaction in clinical trial design.

In time-to-event contexts, we are routinely concerned about events that prevent the existence of measurements, with the question of whether they affect the interpretation or not often a separate and disputable question. The estimands guidance's glossary defines intercurrent events in a way that might address this concern, as "events occurring after treatment initiation that affect either the interpretation or the existence of the measurements associated with the clinical question of interest." The guidance body, however, provides a clarifying example suggesting that the definition of intercurrent event should be limited to events that effect the interpretation, and events that affect the existence of assessments without affecting the interpretation should instead be regarded as missing data. The example showing this is study withdrawal, as distinct from withdrawal from treatment. According to the guidance, while withdrawal from treatment should be considered an intercurrent event, withdrawal from the study as a whole should be handled as missing data.[1]

The estimands guidance also introduces the concept of a *terminal* event. Terminal events prevent the possibility of subsequent measurement. "For terminal events such as death, the variable cannot be measured after the intercurrent event, but neither should these data generally be regarded as missing." There are two examples given in the guidance, death and leg amputation.[1] These examples clarify that terminal events physically prevent subsequent measurement, for any estimand in any study.

We have previously described how censoring and censoring mechanisms for time-to-event endpoints can be embedded into this terminology.[2] In this paper, however, we argue that this terminology may sometimes not be quite adequate to address the needs of time-to-event estimands in the assumption-breaking clinical contexts that oncology sometimes deals with. In time-to-event estimands generally, we need to deal with events that result in not including subsequent measurements in the analysis in a way that is particular to the context of the clinical trial, even if further measurements might be possible in a different study, separately from the question of whether they affect the interpretation of the results. We believe the special clinical context of oncology warrants some supplemental concepts, and care in using the ICH E9 (R1) concepts.

**Oncology Clinical Context**

Oncology has some clinical characteristics that have resulted in therapeutic area-specific endpoints and analysis features such as those previously discussed for censoring and censoring mechanisms.

*High mortality*: One feature pervasive in oncology is the high mortality and short lifespans of patients with advanced cancers. This feature generally precludes a number of analytical methods that might be appropriate in other therapeutic areas. In time-to-event studies for cardiovascular disease, for example, it might be reasonable to assess cause-specific death. Non-cardiac-related deaths are a substantial fraction of all deaths. They can in many cases be objectively distinguished from cardiac-related deaths with reasonable reliability, and the causes might reasonably be considered independent. This situation, however, is rarely the case in late-stage oncology. Cancer-related mortality dominates to the point where the fraction of genuinely non-cancer-related deaths is often negligible, and often few deaths can be clearly adjudicated as having no relationship at all to the cancer. For this reason, assumptions of causal independence that may be reasonable in other therapeutic areas may be less reasonable in oncology or even irrelevant, and cause-specific hazards methods based on these assumptions may be less appropriate.

*Discreteness*: A second feature of oncology trials is the discreteness of many assessments. Evaluation of disease response and progression in particular, critical to many oncology estimands, depends on radiological tumor assessments, labs, physical examinations, and other assessments involving direct contact with the patient. The radioactivity involved in radiological assessments creates risks for patients, and hence cannot ethically be done too frequently. Lab-based and various other assessments can also occur only at clinic visits. Although technology is permitting more data to be assessed outside of clinic visits and with greater frequency, assessment discreteness remains a feature of oncology clinical trials for the foreseeable future. Because time-to-event methods generally assume continuous assessment, the discreteness ubiquitous in oncology often requires special handling.

*Assessments connected to treatment*: An issue closely related to the discreteness of assessments is the fact that dosing in oncology commonly has to occur at the clinic (e.g. intravenously), and is often relatively infrequent, sometimes weekly or monthly. Dosing regimens in oncology can be complex, are often based on drug combinations and may require frequent adjustments based on clinical decisions. This creates a practical need to align clinic visits with dosing schedules. Different treatment arms may have different dosing schedules. Treating physicians generally need safety and disease status data to make dosing and treatment decisions, and patients can only be asked to come to the clinic so often. The practical need to tie assessments to clinic visits, however, creates potential for confounding results. As an example, if safety issues result in delays in treatment in one of the arms, then delays in tumor assessments coinciding with the delayed dosing may result in the less-safe treatment appearing to have a longer time to progression, and hence appearing to be more efficacious.

*Uncertainty of event occurrence and timing*: There is often a gap between the time when assessments first occur and the time when they are interpreted and action is taken on them. The existence of a gap between the two means that other events can occur in-between. Central review and adjudication may result in a further extension of the gap. The investigator interpretation of the results is often used for patient management decisions, such as treatment withdrawal, and may be different from the central review and adjudicated interpretation, which is often preferred in the statistical analysis. A central review board evaluating assessments at the end of the study can, for example, reach a different result from the one reached by the investigator during the study.

*Long-term outcomes*: Many effective anti-cancer therapies do not cure but improve long-term outcomes like disease control and survival. Efficacy assessments require long-term follow-up to assess these outcomes. Clinical trials may last several years. Over this long-time period it is not possible to control external impacts. The therapeutic landscape, standard of care, and patient preferences may all change while the study is ongoing. Patients are undergoing existential challenges; they may experience further comorbidities; their priorities in life may be changing; their motivation to participate in the trial may also change. Patient withdrawal or withdrawal by caregivers is not uncommon in late-stage disease. The greater length of follow-up alone increases the chance of experiencing intercurrent events and/or events preventing further follow-up.

*Scheduling heterogeneity*: Safety-related treatment needs, patient issues, and other factors can result in considerable variability in the time that a given treatment, clinic visit, or assessment actually occurs. Conformance to a precise schedule often cannot be assumed. Assessments are not merely discrete. They are often irregularly discrete.

*Disease heterogeneity*: The causes of cancer are less well-known than for many other diseases. Cancer taxonomy is based largely by site, morphological characteristics, and symptoms rather than underlying causes. Diseases classified as the same cancer based on presenting characteristics might have different causes entirely. While biomarkers and prognostic and predictive factors exist in many cases, there are cases where they do not, and in most cases they are only approximations with limited reliability. Fractional response rates are nearly ubiquitous in oncology clinical trials. Investigators are often unable to reliably limit a treatment population to benefiters at the beginning of a trial, or reliably predict which patients will benefit from the treatment. For this reason, some of the assumptions common in causal inference methodologies, such as no unmeasured confounders, maybe more problematic in the oncology context than they might be in therapeutic areas with more thoroughly studied disease etiologies and treatment mechanisms of action.

*Population heterogeneity*: Study populations are often heterogeneous and treatment effects highly variable. The reasons for disparity of outcomes are often unknown. As discussed below, this heterogeneity of population can confound causal interpretations of common time-to-event estimators and requires care in establishing causal estimands.

*Treatment heterogeneity*: Different local treatment policies may impact the treatment of patients. Even when trial protocols stipulate a standardized treatment strategy, institutional policies or collaborative practice might influence treatment decisions or supportive care.[3] In the same context, oncology treatments often do not follow a fixed treatment regimen. Depending on the disease status, a decision might be taken to start a different treatment. For instance, younger patients with acute myeloid leukemia who achieve remission might receive a stem cell transplant which may be curative. As a consequence, treatment policies and strategies might lead to high variability, have impact on follow-up, and may confound results.

*Non-proportional hazards*: There is an increasing tendency for treatment mechanisms of action to result in nonproportional hazards. Delayed effects and other changes in hazards over time are common in immuno-oncology in particular, but can occur in other contexts as well. Nonproportional hazards can also confound causal interpretations of common time-to-event estimators. In addition, they can make the results dependent on censoring patterns, such as the point at which the study is stopped.

**The impact of non-proportional hazards and heterogeneity on causal estimands**
Two issues mentioned above – the heterogeneity of oncology populations, and the observance of non-proportional hazards in cancer trials – can be particularly problematic in the context of time-to-event analysis. Aalen et al. pointed out that the hazard ratio from a Cox proportional hazards model does not provide a causal prediction of treatment effect given a heterogeneous population.[4] Randomization ensures that at baseline, confounders are reasonably balanced between the treatment arms. However, this balance is not maintained over the course of the trial, because the risk sets at subsequent timepoints become increasingly weighted with lower-risk patients as higher-risk patients experience events. For predictive characteristics, this introduction of imbalance as the trial progresses tends to change subsequent efficacy, which confounds the causality of the treatment effect. The proportional hazards assumption is not maintained, and a Cox hazard ratio is not a causal estimand.[4]

A similar issue occurs when treatment effect hazards are non-proportional, such as when a treatment has a delayed effect. Changes in hazards over time again confound the causality of the treatment effect, violate the proportional hazards assumption, and again result in a Cox hazard ratio not being a causal estimand. As an example, consider a two-compartment fixed delayed effect context. So long as all patients are followed past the delay point, patients censored at study termination have the same future hazards as patients followed longer, and hence censoring introduced by study termination is not informative. Nonetheless, because past hazards (e.g. hazards prior to the delay point) do not predict future hazards, a censored Cox, log-rank, or Kaplan-Meier analysis will produce different results from one in which all patients are followed until event, and hence the censoring introduced by early study termination can be said to introduce bias. Power of Cox models and log-rank tests becomes non-monotone, and results become dependent not just on the point of study termination but on accrual patterns. The estimands guidance presents patient withdrawal from the study, and termination of the study as a whole, as an example of things that "need to be addressed as a missing data problem in the statistical analysis" and do not affect the underlying estimands or interpretation of the results. This example illustrates that in the context of non-proportional hazards, this statement may no longer be appropriate. Termination of the study can affect the estimand and results interpretation.

The special context of oncology, with a variety of clinical features inconsistent with traditionally-made standard assumptions, requires care in formulating and interpreting estimands at the design stage, and also care in clearly understanding and interpreting the available strategies.

**The impact of discreteness and variability of assessment timing**
Standard time-to-event analysis methods assume continuous assessment. The discreteness of assessments in oncology can confound time-to event estimands that use right censoring and assume continuous time.[5,6]

Assessment timing in oncology trials can have significant variation. Dosing and safety assessments often occur more frequently than tumor assessments. Survival is generally assessed daily and can be generally assumed continuous. Accordingly, not just discreteness but differences in schedules need to be taken into account. Progression-free survival is a composite of discretely and continuously assessed elements, progression and survival respectively. Clinical deterioration may result in unscheduled tumor assessments. And safety assessments that may determine treatment withdrawal or otherwise result in the observation of intercurrent events may occur on a separate schedule from tumor assessments. In addition, individual assessments can be subject to substantial variation. Safety issues can delay dosing which in turn delays assessments. Individual scheduling issues may result in variation of days or weeks around target.

The discreteness of assessments inherent in oncology clinical trials might tend to suggest use of interval censoring to account for the impact of the assessment schedule on both the estimand and on estimator operating characteristics. The difficulty is that nonparametric interval censoring methods so far have required not just assessment discretion in individual patients, but reasonably uniform discreteness for the study sample as a whole, with patients generally adhering to an approximately common schedule.[7] Differences in scheduling and variation in timing in practical oncology trials mean that in practice, events can occur and/or assessments end at almost any time, and a specific pattern of discreteness cannot be relied on to occur sufficiently consistently. In group sequential designs, assessment timing irregularities can violate the independence assumptions underlying the use of standard alpha spending functions. Nonparametric interval censoring methods can be sensitive to irregularities and informativeness in interval lengths. The simulations presented by Sun et al.[7] suggest that when regular discreteness can be assumed, and conventional parametric distributions (e.g. exponential, Weibull) are simulated, degradation of testing operating characteristics vs. assuming right censoring is not large as long as the assessment interval remains a fraction of the median survival time, and there are no systemic assessment delays affecting one arm but not the other. While the ability to assume parametric distributions would permit more feasible methods, novel immuno-oncology compound survival distributions in particular, among others, seem to depart from parametric assumptions, making overall inferential validity (as distinct from optimizing operating characteristics) dependent on assumptions a risk.

As a result of these and similar considerations, the combination of complexity due to delays and scheduling variation with relatively small effect on testing operating characteristics has meant that interval censoring was not established as a method of analysis in oncology clinical trials. We agree and accordingly focus on conventional methods with point rather than interval censoring.

Withdrawal
Withdrawal from assessments prevents additional assessments from being collected. It is not in itself an intercurrent event, yet it can have an impact on the results. When withdrawal occurs for a reason potentially related to the treatment efficacy or safety, such as due to radiological or clinical progression, treatment failure, adverse event, and so forth, an intercurrent event can be identified.  Dating withdrawal and identifying the reason for it requires special care in oncology settings. The discrete nature of assessments complicates matters. There is often a time interval in-between. Differences in assessment scheduling for different kinds of assessments may mean that potential intercurrent events can be assessed between assessments of the underlying event of interest. Further, patients are withdrawn following interpretation, but withdrawal-precipitating events are traditionally dated at the date of assessment. Gaps between detection and interpretation can result in an event of interest occurring between the two, potentially confounding the chain of causality.

In the estimands framework, if the withdrawal was precipitated by an underlying event potentially related to treatment efficacy or safety, this precipitating event represents an intercurrent event.

This complicated state of affairs has not been adequately dealt with in the literature. Careful attention is needed to the point at which a patient exits the risk set.

In addition, censoring is only one way of handling events precipitating withdrawal, and specifically assumes such events are not informative. We think there is merit in a terminology that would apply more generally, both to cases where censoring is appropriate and to intercurrent events where censoring is not appropriate.

**Occlusion**

To help address these concerns, we propose the concept of occlusion. In our first paper,[2] we briefly introduced the concept of occluding events, with occlusion defined as any loss to further follow-up and/or removal of further collected data from analysis without having experienced the event of interest. We believe this is the broadest generalization appropriate to a time-to-event context. In particular, occlusion occurs at the point the patient leaves the risk set. As we discuss in more detail below, the discrete and varied nature of assessments and decision-making in oncology means that this point may be different, for the same underlying data, for different estimands.

Under this conceptualization, occlusion represents generally the act preventing observation of or resulting in discarding data. Censoring is a kind of occlusion, including both censoring as the result of the pre-specified end of the study, and censoring which results in discarding data at analysis time that was actually observed and collected in the study. But occlusion need not be handled only through censoring. It could be handled through, for example, a competing risk event, composition, or a (causal) hypothetical method, all of which also result in the analysis not looking further than the occluding event. From this perspective, censoring is a particular technique that arises in analysis contexts, while occlusion is a considerably broader concept (Figure 1).

Occlusion is also a broader concept than a terminal event as defined in ICH E9 (R1).[1] Terminality is an objective property of an event which renders further observation physically impossible. If an event is terminal, it is impossible to devise a study that can look beyond it. Indeed there is no meaningful clinical question regarding the treatment effect that manifests after a terminal event. Occlusion, however, is context-specific. It can arise from a decision not to observe further, or even a decision to discard data already observed. As an example, if a study uses a treatment policy strategy to address subsequent therapy but then includes a supplemental analysis with a hypothetical strategy, subsequent therapy will be an occluding event for the hypothetical estimand. But it will not be an occluding event for the treatment policy estimand in the same study. Occlusion may also be a consequence of study procedures. The primary estimand might not require data collection beyond a certain event. Since the primary estimand might control data collection in general, this event might become an occluding event for secondary estimands.

With the concept of occlusion in hand, we can extend the concepts of noumenal and phenomenal censoring introduced in our first paper[2] to noumenal and phenomenal occlusion. We previously defined noumenal censoring as an estimation method to address an intercurrent event which influences the clinical question itself (i.e., defines the pre-specified estimand), distinguishing it from phenomenal censoring that addresses administratively missing information (i.e., defines missing data handling in the estimate). We borrowed the terms noumenal and phenomenal from the epistemology of Kant, with *noumenon* the unobservable thing-in-itself and *phenomenon* what appears to the senses.[8,9] This terminology translates well into the conceptual framework of the estimands language, which broadens the concept of a population of inference to include hypothetical populations. The estimands framework addresses the fact that the clinical environment of interest may sometimes be so different from the clinical trial environment as to render it not directly observable. In our formulation noumenon describes the estimand, the environment of inference; phenomenon describes what is observed in the trial. And unlike a noumenon in the Kantian context, where a noumenon remains an external thing that is not directly observable, a noumenon in an estimands context includes not just future but counterfactual frames of reference.

**Noumenal occlusion**
When occlusion occurs as a systematic part of the estimand definition or study design, e.g. when the withdrawal criteria or visit schedule stop collecting assessments at an event or the censoring table censors for it, we call this occlusion noumenal. Noumenal occlusion is a property of the estimand. It represents a design decision that we do not wish to look beyond the occluding event. It should reflect an explicit definition of the relevant clinical question based on a careful discussion among the design team that explains why looking beyond the occluding event is clinically irrelevant, infeasible, or inappropriate. This may be because the occluding event is considered related enough to the event of interest that it can be subsumed into it (composite strategy); because we are interested in what would have happened if the occluding event had not occurred (hypothetical strategy); or because we are not interested in what happened after the occluding event (while-on-treatment or while-prior-to-occlusion strategy).

When noumenal occlusion is handled through censoring, noumenal censoring results. Examples of noumenal censoring include a censoring table that always censors for subsequent therapy. But noumenal occlusion also occurs when e.g. a while-alive strategy is implemented by making death a competing risk event, a composite strategy is implemented by making death a composite event, or a hypothetical strategy handles treatment switching using a causal inference technique. All design and analysis choices which result in systematically not looking beyond an event represent noumenal occlusion.

In our first paper, we discussed implicit noumenal censoring.[2] This occurs when a study systematically stops assessments at an event, for example when under a study design clinic visits end at radiological progression, but this event is not acknowledged in the intercurrent events strategy or censoring table for other secondary endpoints that are also based on clinic visits. Implicit noumenal censoring often occurs when secondary estimands must be fit into a study whose visit schedule and withdrawal criteria are determined by the needs of patient management and the primary estimand. It is always a kind of implicit noumenal occlusion. In principle, implicit noumenal occlusion could be implemented by some means other than censoring, perhaps by some method yet to be devised. We struggled, however, to come up with a current example of implicit noumenal occlusion by means other than censoring. Current means of handling occlusion other than censoring, such as composition, competing risks, and causal inference methods, require explicit specification of how these events are handled. Censoring in contrast is used as a catch-all method for handling termination of assessments for unspecified reasons.

An advantage of the concept of occlusion is that it applies across multiple strategies for addressing intercurrent events. Hypothetical, while-on-treatment, and composite strategies all use noumenal occlusion to address an intercurrent event. Treatment policy strategies and principal component strategies do not use occlusion. An event that would be noumenally occluding under a hypothetical, while-on-treatment, or composite strategy would not be noumenally occluding under a treatment policy or principal component strategy.

**Phenomenal occlusion**
Phenomenal occlusion is occlusion that arises under study conditions, and not from a systematic element of the estimand and thus, of the study design. Examples include individual-patient withdrawal decisions, equipment failures, and site closures.

Phenomenal occlusion may be informative. Individual patients may withdraw, for example, for reasons entirely related to their future prognosis. Even when the study team has done its best to identify likely intercurrent events, collect data supporting a determination if occlusion is treatment-related, etc., some confounding of the study is perhaps inevitable. However, when the team has made efforts to pre-specify and collect data on likely intercurrent events, and conducted appropriate sensitivity analyses addressing informativity, phenomenal occlusion can then be assumed uninformative and handled through censoring.

In practice, other ways of handling occlusion (competing risk event, composition, causal hypothetical, etc.) require explicitly specifying the event and handling all events of that type the same way, while unspecified and unforeseen occluding events are handled through censoring as a catch-all. For this reason, phenomenal occlusion generally involves censoring.

[Insert Figure 1 here]

**Additional details on strategies**

The discussion of intercurrent event strategies in our first paper[2] is assumed. We provide additional comments on specific strategies.

**Treatment policy strategy**

The treatment policy strategy, in its original context, focused specifically on the intercurrent event of treatment switching. In that context, it compared the randomized treatments plus all subsequent therapies. The 2007 FDA Cancer Endpoint Guidance,[10] in effect until replaced by the 2018 Cancer Endpoint Guidance,[11] did not use this approach. It assumed that the treatment effect of interest in a clinical trial is only the effect of the original randomized treatment, and hence that subsequent therapy confounds this effect. Accordingly, it presented as an advantage of PFS, that "its determination is not confounded by subsequent therapy."[10] This interpretation of treatment effect led to the 2007 guidance's recommendation to censor for subsequent therapy.

Carroll[12] and Fleming et al.[13] proposed a different characterization of the treatment effect which Carroll called "treatment policy."[12] In this view, the effect of the original randomized treatment regimen alone does not provide an adequate assessment of the future consequences of the treatment course. As Fleming et al. explained:

> Often, an important component of the effect of an experimental agent is its impact on the need for supportive or subsequent care, the ability to receive it, and what effect it might provide. Therefore, a weakness rather than strength of PFS (relative to OS) is that it does not incorporate the real world long-term impact of the experimental regimen between the times of disease progression and death. The trial should be designed to ensure these important longer effects can be properly estimated.[13]

In this strategy, subsequent therapy and its consequences are included in the research question. With this defined research question, differences in subsequent therapy behavior between the treatment arms do not result in bias but are part of the effect to be estimated.

The ICH E9 (R1) guidance adapted the Carroll[12] and Fleming et al.[13] concept of a treatment policy strategy, and generalized it to cover any intercurrent event. Under the guidance's formulation, "the occurrence of the intercurrent event is considered irrelevant to the treatment effect of interest: the value for the variable of interest is used regardless of whether or not the intercurrent event occurs."[1] In our view a treatment policy strategy as adapted by the guidance has two implicit additional features. First, it requires systematically following patients through and beyond the intercurrent event until the clinical event of interest is observed. Second, it assumes that study conditions are predictive of real-world clinical practice with respect to that event.

From this point of view, pervasive occlusion, even the end of the study (as we discuss below), can defeat a treatment policy strategy.

A potential source of confusion in the estimands guidance is that several of the key terms, including "treatment policy," were named in the context of a particular intercurrent event, but are general terms that can be used for any kind of intercurrent events. For example, events like radiological progression are generally predictive of future mortality and hence are intercurrent events with respect to survival. Survival trials typically follow patients through and beyond such events, and ignore them in the OS analysis. They accordingly apply a treatment policy strategy for these other intercurrent events. The interpretation is analogous; overall survival includes survival with respect to the complete patient history both before and subsequent to them.

**While-on-treatment/While prior to occlusion strategies**

A while-on-treatment strategy is only concerned with what happens up to the time of the intercurrent event. We have found in our experience that the term "while-on-treatment strategy" may be somewhat confusingly named. The term is tied to the context of treatment withdrawal and treatment switching. But the strategy can be applied to any kind of intercurrent event.

In our practical experience applying the estimands framework, we have found that the term "while on treatment" tends to cause confusion when it is proposed to apply it to an intercurrent event other than end of treatment, and it has often been necessary to explain that such an application is possible. Based on our experience that this term has been confusing in practice, we suggest the term "while prior to occlusion" (or "while prior to event") as a general name for the strategy. Formulating each estimand, the strategy could then be named for the specific occluding event of interest. For example, "while alive" for death, "while progression-free" for progression, "while on treatment" for treatment withdrawal (only), etc. It should be noted that while on treatment/while prior to occlusion strategies always involve noumenal occlusion, as anything after the occluding event, being irrelevant to the scientific question of interest, is always removed from the noumenon.

In our first paper,[2] we discussed two basic models used to implement a while-prior-to-occlusion strategy for time-to-event estimands, the cause-specific hazards model and the subdistribution hazards model.[14,15] As discussed in the clinical context section above, we suggest causal independence can rarely be reliably established in oncology. Accordingly, we do not recommend cause-specific event time-to-event estimands except in unusual situations, where events of interest can be reliably assumed to be independent of occluding intercurrent events.

Some additional comments should be noted about the use of while-prior-to-occlusion strategies in particular contexts.

**Time-to-event safety estimands**

In most regulatory oncology trials, general follow-up for safety is limited to a "treatment emergent" period from start of treatment to a short time following last treatment, commonly 30 days. The reason for this traditional approach is an assumption that safety events are only caused by treatment while the treatment is still circulating within the patient, and hence causality ends after a certain number of pharmacokinetic half-lives from treatment end. The follow-up period may be longer for classes of therapy (e.g. immunotherapy) with delayed safety effects.

When developing time-to-event safety estimands involving treatment-emergent events or other safety events implemented with time-limited follow-up, a competing-risk while-on-treatment strategy may be more appropriate than traditional censoring. Patients in oncology studies tend to end treatment at different times. When this occurs, the rationale for limiting the treatment-emergent period results in a violation of the non-informativity assumption. Hazards during the treatment-emergent period are different from hazards afterwards. The hazards of patients after their treatment-emergent cutoff cannot be validly predicted by the hazards of patients who are still in the treatment-emergent period at the same time calculated from treatment start. The competing-risk approach to a while-treatment-emergent strategy, which looks only at what happens during the treatment-emergent period, does not have this difficulty, and does not require assuming non-informativity

As discussed in our first paper,[2] a difficulty with a while-on-treatment approach in a context where causes are dependent is that Fine-Gray subdistribution hazards model, used in modelling and testing, has dependent hazards issues which do not result in a causal estimand or a clear clinical interpretation. The descriptive Cumulative Incidence Function (CIF) does not have this defect. Because safety estimands often only require only a descriptive approach, it may be appropriate to consider in this context.

**Patient-reported outcomes (PROs)**

When PROs are collected at clinic visits, assessments normally end at events that end clinic visits. Active follow-up with clinic visits often ends either shortly after treatment withdrawal or at radiological progression. Accordingly, in this situation, events like the end of the treatment period or radiological progression represent noumenal occluding events for PROs, events that would generally be related to PRO outcomes and hence intercurrent. Even for withdrawal based on patient decision, the reasons to withdraw may be related to perceived outcomes and quality of life.

Accordingly, for withdrawal-inducing events posited as highly related to PRO outcomes, a while-prior-to-occlusion strategy might be considered. However, as discussed above, we do not generally recommend independent-causes models in oncology, and for a competing risks approach we suggest descriptive approaches such as the CIF due to problems with interpreting the Fine-Gray subdistribution hazards model. A descriptive approach may be more restricting in a PRO context because modelling and testing are often desired.

Technology is increasingly permitting PROs to be assessed using portable electronic devices which do not require clinic visits and can permit assessments to continue after clinic visits end. Study teams concerned about the possible intercurrence of occluding events on PRO results should explore technology needed to enable PRO assessments to continue. For example, in a context where patients stop therapy and most clinic visits early and anticipate an improved PFS in conjunction with a treatment-free period with no side effects, the relevant clinical question might target quality of life regardless of treatment discontinuation, making a treatment policy strategy more appropriate to the clinical question of real interest. In such cases, ways to facilitate assessing PROs beyond the end of therapy and without needing clinic visits would be particularly helpful and should be sought out.

**The end of the trial**

As discussed above, ICH E9 (R1) presents the administrative end of a trial as an example of missing data, to be handled through the statistical analysis method, with no potential to affect the choice of estimand or influence the result.[1] This assumption, however, does not necessarily hold in the context of non-proportional hazards, which can arise both because of the heterogeneity of populations common in oncology, and the tendency towards delayed treatment effects and variable hazards present caused by treatment mechanisms of action. Non-proportional hazards occur in a number of oncology contexts and should be considered as a possibility in any major time-to-event oncology study.

Because of the tendency for results to depend on the duration of the trial, we recommend interpreting the end of the trial as inducing a "while during trial" strategy, rather than as representing pure missing data. Doing so clarifies, like any "while prior to occlusion" strategy, that the research question involved is only concerned with outcomes occurring during the trial. With this interpretation, the results are unbiased: they answer exactly the research question posed. In this context, independent-causes censoring is generally appropriate. As a consequence, a reasonably well-designed trial is needed and care will have been taken to follow all or nearly all patients past the delay point, and censoring for end of trial will be non-informative. The analytical method therefore remains exactly the same as for a treatment-policy strategy, in which the end of the study is interpreted (per the Guidance) as pure missing data with no effect on the results. Nonetheless, interpretation as a "while-during-trial" strategy is different from interpretation as an intercurrent treatment-policy strategy.

A problem with simply interpreting a standard analysis as an independent-causes while-during trial strategy, however, is that while the results become interpretable, and the interpretation is valid, nonetheless the interpretation is not reproducible. Studies inevitably end at different times and have different accrual patterns. Accordingly, simply interpreting standard study results as an analog of a while-on-treatment strategy renders the interpretation valid, but valid only to that specific study.

We suggest either an estimation approach that does not assume proportional hazards, or one that more directly addresses the time-limited nature of a while-during-study estimand.

Aalen et al.[4] suggested accelerated failure time or additive hazards models as an alternative not dependent on the non-proportional hazards assumption. We believe such an approach would be needed to permit a treatment policy estimand in a non-proportional hazards context.

Within the context of an underlying model that assumes proportional hazards, the restricted mean survival time (RMST) has been proposed to address causality under a breakdown of the proportional hazards assumption.[16] We would note that in a non-proportional hazards context, time-restricted estimators like RMST remain dependent on the accrual pattern, and hence do not completely solve the reproducibility problem. Nonetheless, we believe they are preferable to unrestricted estimators in this context. Time-restricted estimands with explicit and comparable timing clarify the restricted nature of the conclusions supportable from the study, and in our view improve interpretablility. Using time-restricted estimands to address non-proportional hazards illustrates that in this context, the appropriate interpretation is indeed an analog of a while-on-treatment/while-prior-to-occlusion strategy, and not a treatment policy estimand.

**Hypothetical strategies**

In our first paper,[2] we discussed a particular context in which a hypothetical strategy might be considered, a situation where the trial context induces patient behavior that would not occur in an ordinary clinical setting. In this context, the clinical trial context is itself counterfactual to what we expect to occur in clinical practice. For this reason, a counterfactual causal-inference strategy, despite strong assumptions and doubts about its reliability in providing answers, at least has the benefit of addressing the right question. See generally Manitz et al.[17]

We would like to make some additional comments on this issue. Deming distinguished between enumerative studies, in which inference is based on sampling from a frame with error assessable based on a sampling error alone, and analytic studies, which lack a sampling frame and hence whose reliability cannot be based solely on sampling error.[18] Deming pointed out that any attempt to predict the future is necessarily analytic in character, as the future has no frame from which to sample,[19] and stated that management is prediction.[20]

Clinical trials are attempts to predict the future, and registrational trials attempt to predict outcomes following approval. While frequentist inference today, especially in clinical trials, is generally based on randomization inference rather than sampling inference,[21] nonetheless the enumerativity assumption fundamental to statistical inference – that statistical error approximately assesses overall reliability error – includes an assumption of environmental stability, that the future will resemble the past. A key element of the estimands guidance is an acknowledgment that this assumption does not always hold in the context of clinical trials. It is not just that once randomization has completed, what follows is in part an observational study with opportunities for intercurrent events and occlusion to confound inference. The very fundamentals of the trial itself – the processes of randomization and placebo control which statistical inference its rigor – can induce artificial behavior in patients that tends to break the enumerativity assumption.

We do not suggest that current causal-inference hypothetical estimands, while an attempt at a solution, are the ideal solution, only that they may sometimes be the best currently available under the circumstances. As noted in our first paper,[2] their benefit is that they attempt to address a relevant clinical question. We would note that being able to make reasonably reliable decisions in a context where pure statistical inference is regularly confounded, in this context as in others, requires combining statistical knowledge with insight into underlying processes, complex behavior, the effect of system, and interaction.

**Withdrawal reconsidered**

As discussed above, withdrawal from assessments can represent something of a special case. It is an inherently occluding event and can have an impact on the study results, but it is not, by itself, an intercurrent event. In this section, we discuss considerations for documenting and determining the timing of occlusion in the context of withdrawal.

The special context of oncology results not just in discrete assessments with potential for events to take place between assessments, but a frequent need for time to interpret and adjudicate assessments that can result in a lag not just between events and their physical detection, but in lags between physical detection, interpretation, and action. Because of this potential for time lag, a last assessment could take place between a precipitating event and the withdrawal, and an event of interest could take place during this time as well. Accordingly, there are five kinds of events to consider: (1) the precipitating intercurrent event, (2) first assessment of the intercurrent event; (2) additional assessment(s), (3) the event of interest, (4) identification of the intercurrent event, and (5) the actual withdrawal.

As an example, consider a study with weekly dosing, tumor assessments every 6 weeks, and a time to event variable based on PROs assessed at e.g. every dosing visit, or a daily patient diary, with patients withdrawing from assessments at progression (Figure 2). In this case, multiple PRO assessments could occur between an unknown actual progression and its detection (D6), and further PRO assessments could occur between the tumor scans and the subsequent determination that the patient has progressed and the actual withdrawal. If progression is re-evaluated by a central radiology review at the end of the study, the central assessment could differ from the investigator assessment on which withdrawal decisions were based.

[Insert Figure 2 here]

Special care should be taken to determine:

- The time at which the patient leaves the risk set. This could occur at the point of first detection of the triggering event (D6), the point of actual withdrawal (V), or the point of last assessment if in between (D7); and
- What to do if the event of interest (or an assessment) occurs between the triggering event and actual withdrawal.

We suggest two principles to help guide the decisions involved. First, causality should go forward in time, so a decision whether a triggering event occurred or not cannot be based on what happened afterwards. This would suggest that if the occlusion is dated at the triggering event, any subsequent events would be discarded. Second, potential for manipulation should be avoided. If the investigator is allowed to discard observed events based on a subjective determination that the patient would have been withdrawn had the event of interest not occurred shortly afterwards, a potential for manipulation occurs. For example, if a patient dies while on treatment, it is quite likely that close examination of the patient journey would identify an event triggering mandatory withdrawal occurring shortly before the death. The ability to identify such an event and throw out the death based on investigations and decisions made after the death occurs, knowing about the death, could be subject to unconscious biases. Accordingly, when the determination whether a triggering event occurred or not is subject to judgment with potential for unconscious biases, this consideration would suggest it might be better to keep the patient in the risk set until, and date the occlusion at, the actual withdrawal (or last assessment prior to withdrawal) rather than the triggering event.

Applying these questions to the strategies, we would note that these ideal principals may require careful thought in particular cases and may have some potential for compromise. Assessment of withdrawal-precipitating events in cancer trials, including treatment withdrawal decisions, progression determinations, and others, often involves an element of subjective judgment, so that manipulation-avoidance concerns would tend to favor including all assessments and events observed up to the point of actual withdrawal. However, the research question in the hypothetical and while-prior-to-occlusion strategies is only concerned with what happened up to the occluding intercurrent event, and the best answer to that question without reliability considerations would tend to discard what occurred after an occluding intercurrent event. We suggest a balance between these considerations.  We do not believe there are hard-and-fast one-size-fits all answers, and these issues deserve additional attention from the research community There may be a degree of acceptable subjectivity.

One suggestion we would make is that if there is an overall study approach to the question of subjectiveness of an event this might be appropriate to apply the same approach to the subjectivity of corresponding withdrawal decisions based on the event. For example, for a study that requires central radiological review of progression based on a determination that investigator view is insufficiently objective, it might be appropriate to consider investigator-determined progression a subjective decision for withdrawal purposes. But in a study whose primary analysis is based on investigator determined progression, it might be appropriate to consider investigator-determined progression sufficiently objective in the withdrawal context as well.

**Withdrawal considerations by strategy**

The appropriate approach depends on the strategy used.  The following should be considered:

- When a treatment policy strategy is used for the intercurrent event underlying the withdrawal, consistency with the general approach for a treatment policy strategy would suggest the intercurrent event should be ignored, and the last assessment date on or prior to withdrawal used for phenomenal censoring since this date confirms absence of the event of interest.
- Similarly, if a principal stratum strategy is used, withdrawal-precipitating intercurrent events represent individual exceptions to the principal stratum model, and hence represent phenomenal occlusion. The approach would then be similar to a treatment policy strategy, with the intercurrent event ignored, and the last assessment dated on or prior to withdrawal used for censoring.

- When a composite strategy is used for the intercurrent event, the intercurrent event becomes an event of interest, and everything that occurs afterwards, including the original event of interest if subsequently observed or not, is ignored.
- A hypothetical or while-on-treatment strategy may warrant further inquiry.
    - If withdrawal is triggered without subjective judgment, then it might be appropriate to date the occlusion at the triggering intercurrent event, with any subsequent data ignored.
    - If withdrawal is potentially based on subjective judgment, it might be appropriate to use the date of the actual withdrawal. If the event of interest is observed after the triggering event and before withdrawal could occur, it would then be counted.

For each estimand, the relevant withdrawal is from the specific assessments for that estimand.

**Withdrawal and composite events**

Different components of a composite event can sometimes be assessed on different schedules, and patients can potentially withdraw from different assessments at different times. An important practical implementation question that needs to be determined is when the composite assessment period ends where individual components end at different times. For this reason, defining the last assessment for a composite endpoint requires care.

In its 2007 Cancer Endpoint Guidance, the FDA introduced a "2 missing assessment" rule for PFS by which an event would not be counted if there were at least 2 missing assessments between the last tumor assessment and the event.[10] This rule was a way of implementing the concept that at some point a patient stops being "at risk" for tumor assessments, and unless the patient is at risk for all components of the composition, the patient is not at risk for the composite endpoint.

In the case of noumenal occlusion, patients should be regarded as no longer being at risk for the composition as a whole when they stop being at risk for any noumenally occluded component. As an example, consider a "symptom deterioration free survival" endpoint in which symptoms are assessed at clinic visits while survival assessments continue into long-term follow-up. If clinic visits end at radiological progression, implicit noumenal occlusion occurs. We suggest that the composite estimand as a whole should be regarded as being occluded at radiological progression, and we would suggest considering an appropriate strategy, other than treatment policy, to address this noumenal occlusion. Phenomenal occlusion, however, may permit more flexibility. As an example, for a progression-free survival estimand with the above visit schema, if patients are systematically followed for progression in the course of the study, then isolated cases of patients continuing survival follow-up after ending tumor assessments without progression would not necessarily be inconsistent with a treatment policy strategy for the composite endpoint as a whole.

Accordingly, the concept of when a patient is "at risk" for an event requires special attention. The underlying principle is that if the patient will not be assessed, the patient is not at risk for an event being detected, and hence the relevant question is at what point a patient stops being at risk for being assessed. This approach makes the question dependent on the particular trial design and context.

The situation becomes a bit more complicated when either component of a composition can occur first and have assessments end first, and is further complicated when both components are assessed discretely, and on different schedules. Careful attention is needed to identifying when the patient leaves the risk set.

- In situations where unscheduled assessments are uncommon, the assessments can be assumed discrete. Once an assessment occurs, the event could be assumed undetectable until the next assessment. If assessments for other components of the composition occur between the two, they could be counted.
- In indications where unscheduled tumor assessments are common, the assumption that tumor assessments are continuous is more reasonable, then a new assessment might occur at any time following the last assessment, and the patient would be continually "at risk" for being assessed. The last assessment date of the first component to end assessments would be more reasonable.

Let us revisit our example (Figure 2) with a "symptom deterioration and progression free survival" variable where symptoms are assessed at clinic visits approximately every 2 weeks, tumor assessments for progression are assessed every 6 weeks, and death can be identified to the day.

In this example, consider a patient whose tumor assessments end prior to the end of safety clinic visits, e.g. last tumor assessment is Week 18 and symptom deterioration assessments continue to e.g. Week 30. In this case the first missed assessment would be the Week 24 tumor assessment, and the last prior assessment would be the Week 22 safety assessment.

- If detecting progression through unscheduled tumor assessments is not common, the study team might consider ending assessments for the composite variable there, counting the Week 20 and Week 22 safety assessments and any event observed. If discreteness of assessments can be assumed, the patient can be assumed not to receive another tumor assessment until Week 24, and accordingly the patient is fully assessed through Week 22, not just through Week 18.
- If detecting progression through unscheduled tumor assessments is common, the study team might regard the patient as ending assessments for all purposes at Week 18.

Now consider a patient whose tumor assessments end after the end of safety clinic visits, e.g., symptom deterioration assessments end at Week 26 but tumor assessments end at Week 30.

- In this case the first missing assessment is the Week 28 symptom deterioration assessment, and the last prior assessment is the Week 26 symptom deterioration assessment. We would suggest it would be reasonable to end assessments at Week 26 regardless of whether unscheduled assessments were common or not.

These considerations apply particularly to noumenal occlusions, where assessments are systematically ended. Phenomenal occlusion, isolated or occasional ending of assessments, may offer more basis for flexibility.

### Additional considerations for sensitivity analyses

**Sensitivity analyses for timing of assessments**

Interval censoring methods could potentially be done as a sensitivity analysis for the assumption of approximately continuous censoring. Sun (2013) discuss ways of grouping data into "bins" to reduce the impact of assessment timing variation. A simpler sensitivity analysis to assess the potential impact of using right censoring on the results would date both events and censoring at, respectively, the day after the last prior assessment (left censoring, at the beginning of the interval) and also at the midpoint between the last prior assessment and the event/censoring point (middle of the interval. This approach could also be used as an alternative to the 2007 FDA Cancer Endpoint Guidance[10] approach of censoring variables with 2 or more intermediate missing assessments at the last contiguous assessment. A second approach to assess the impact of potentially informative departures from target assessment times on the results would be to assess the distribution of departure from target assessment time in each treatment arm, and to perform an analysis with assessment times imputed as the target time per the protocol assessment schedule.

**Sensitivity analyses for withdrawal**

In cases where a withdrawal-precipitating intercurrent event can occur prior to withdrawal, we recommend sensitivity analyses dating occlusion (1) at the time of the intercurrent event, discarding any information occurring afterwards, and (2) at the time of actual withdrawal, to evaluate the effect of the timing of the interval.

**Discussion and conclusion**

In our first paper,[2] we introduced the concept of noumenal and phenomenal censoring, and briefly introduced the concept of occlusion. This paper describes the concept of occlusion in greater detail. We define *occlusion* as any event resulting in stopping further observation and/or not including subsequent data. Under this approach, occlusion represents the actual stopping of assessments or discarding of data, with no assumptions made, beginning at the point the patient leaves the risk set, while censoring is a technique of addressing occlusion which requires assumptions to be valid. We propose this concept because of the clinical complexity of oncology studies, where due to heterogeneity, discreteness of assessments, and other factors, neither point censoring nor interval censoring assumptions fully apply.

We adopt a broad conception of occlusion, rendering it relative to choice of study design, estimand, and analysis. This choice gave rise to discussion within the Censoring Mechanisms Subteam. Occlusion could be seen as either an objective property of the event, or one that might depend on study design, but not depend on choice of estimand within a single study. Under this view, occlusion would remain a property of events, at least within the context of a study, and not depend on the research question posed about them or the method of analysis used. This would permit designers to first determine if an event is occluding or not, and then determine the appropriate strategy to deal with it, as they do with intercurrent events per ICH E9 (R1). A definition in which occlusion depends on choice of strategy defeats this order.

The chosen approach permits a clearer and more precise way to handle withdrawal, which depends on choice of estimands. As we have discussed, the point at which the patient leaves or is removed from the risk set depends on exactly what the patient is considered at risk for, and this depends on the choice of estimand. We recommend looking at the estimands choice, study design, and visit schedule as an integrative whole, acknowledging they form a system and affect each other.[2]

Flexibility in choice of estimands is an important value. Different estimands are appropriate to different audiences and purposes. For example, while understanding the complete consequences of an initial treatment assignment decision including all subsequent therapy may often be of interest to scientists and regulators concerned with the effect of treatment policies on populations, patients and their healthcare providers may sometimes be interested in attempting to estimate the effect of a particular treatment decision in isolation rather than in its combined effect together with all subsequent therapy.

The broader definition of occlusion acknowledges this diversity and retains maximum design flexibility, permitting establishing desired estimands up front and then evaluating implementation feasibility. Clinical development planners can choose to evaluate two estimands in two different studies, collecting data past an intercurrent event in one study and stopping at the event in the other. Or they can instead evaluate both estimands in the same study, collecting data past the intercurrent event but knowingly discarding it for the estimand which does not use it.

Because intercurrent events tending to precipitate withdrawal can occur well before actual withdrawal with assessments and even events in between (i.e. withdrawal may be precipitated but fail to materialize), the time of occlusion is sometimes not observable and may have occurred within a wide interval of time, and identification of the appropriate point within that interval requires careful consideration.

In this paper we further develop the concepts of noumenal and phenomenal introduced in our first paper,[2] applying them to occlusion. Noumenal occlusion occurs when data following an event is not observed or excluded by the construction of the estimand and thus, by study design, systematically for patients generally. Noumenal occlusion is implicit if the study design stops assessments at a particular event but this event is not mentioned in the estimand definition or censoring table. Noumenal occlusion is immanent in the estimand definition and can be handled in a variety of ways, not just through censoring, but through for example a competing risk event, composition, or a counterfactual causal inference method. Phenomenal occlusion is occlusion in an individual patient caused by on-study events, such as a decision to withdraw or loss to follow-up for reasons not related to the treatment outcome. While we elaborate the concepts of noumenal, phenomenal and occlusion in the context of time-to-event estimands in oncology clinical trials, we believe that these concepts are also applicable and relevant in non-oncology settings.

We introduce some background information on the oncology context. The terminology introduced in the Estimands Guidance, including intercurrent event, missing data, and terminal event,[1] does not completely address the conceptual needs of time-to-event estimands in the especially complex oncology context. We introduced the concept of occlusion to supplement. We also suggest that the while-on-treatment strategy be renamed so as not to be tied to the specific intercurrent event of withdrawal, clarifying interpretation in a more general context. We propose while-prior-to-occlusion instead.

We propose detailed considerations for handling specific example situations, including safety estimands, PRO estimands, withdrawal from assessments, and the end of the study. We also propose sensitivity analyses to address the interval nature of assessments in oncology trials, and intervals between withdrawal-precipitating intercurrent events and actual withdrawal.

We suggest that consideration of quality management principles and caveats regarding the reliability of statistical inference[20,22] might be useful. As we discussed in our first paper,[2] the research goals and practicalities of clinical trial operations can sometimes require compromising, and components of a trial can sometimes constrain and interfere with each other. Where the needs of one estimand result in ending assessments for another, we recommend making the relevant occlusion explicit, and handling assessment-ending intercurrent events with an appropriate strategy. In general, we recommend carefully defining and prioritizing goals, designing the study with those goals in mind, and making careful choices and compromises so as to optimize the whole. The best solution for the trial as a whole may require a sub-optimal solution for each individual component.

In designing trials, we recommend close cooperation among clinicians, statisticians, and trial management and operations professionals, attempting to understand the overall needs of the study as a team, sometimes extending beyond the boundaries of each individual discipline. Trials can themselves act as a system, interacting and interfering with each other. Although causal-inference statistical adjustment techniques and hypothetical strategies can help address symptoms of this interference, we suggest attempting to address the cause.

David Kerridge characterized Deming's contribution to statistics as understanding when statistical inference works and when it doesn't.[23] The Estimands Guidance brings the need for this understanding to the forefront of applied statistical practice. Really grappling with the assumptions involved in applying statistical methods requires combining theoretical statistical expertise with philosophy of scientific inference, a systems approach, practical knowledge of the clinical trial environment, cooperation with others, and curiosity about the world.

In many ways, this is an exciting time to be a statistician. Understanding the limits of our knowledge helps improve it.

**Data availability statement**

No new data is presented in this manuscript. This manuscript is based solely on previously published results.